\documentclass[10pt, a4paper, twocolumn, twoside]{article} 

\usepackage[english]{babel} 

\usepackage{microtype} 

\usepackage{amsmath,amsfonts,amsthm} 

\usepackage[svgnames]{xcolor} 

\usepackage[small, labelfont=bf, up]{caption} 

\usepackage{booktabs} 

\usepackage{lastpage} 

\usepackage{graphicx} 

\usepackage{subfigure}

\usepackage{enumitem} 
\setlist{noitemsep} 

\usepackage{sectsty} 
\allsectionsfont{\usefont{OT1}{phv}{b}{n}} 

\usepackage{txfonts}

\usepackage{geometry} 

\usepackage[T1]{fontenc} 
\usepackage[utf8]{inputenc} 

\usepackage{XCharter} 

\usepackage{hyperref}

\usepackage{journals}

\usepackage{fancyhdr} 
\newcommand{\shorttitle}[1]{\fancyhead[CE]{\textsl{#1}}}
\newcommand{\shortauthors}[1]{\fancyhead[CO]{\textsl{#1}}}

\fancypagestyle{firstpage}{ 
	\fancyhf{}
        \lhead{\vspace*{0.0em} \textit{\footnotesize 21$^{\rm st}$ European Workshop on White
            Dwarfs \\ Castanheira, Campos, Montgomery, eds. \\[-0.2em]
          July 23--27, 2018, Austin, Texas, USA}}
}

\date{}

\newcommand{\authorstyle}[1]{{\large\usefont{OT1}{phv}{b}{n}\color{DarkRed}#1}} 

\newcommand{\institution}[1]{{\footnotesize\usefont{OT1}{phv}{m}{sl}\color{Black}#1}} 
\newcommand{\email}[1]{{\footnotesize\usefont{OT1}{phv}{m}{sl}\color{Black}#1}} 

\usepackage{titling} 

\newcommand{\HorRule}{\color{DarkGoldenrod}\rule{\linewidth}{1pt}} 

\pretitle{
	\vspace{-30pt} 
	\HorRule\vspace{10pt} 
	\fontsize{16}{12}\usefont{OT1}{phv}{b}{n}\selectfont 
	\color{DarkRed} 
}

\posttitle{\par\vskip 10pt} 

\preauthor{} 

\postauthor{ 
	\vspace{0pt} 
	{\par\HorRule} 
	\vspace{-10pt} 
}

\newcommand{\newabstract}[1]{
    {\section*{Abstract}
    \bfseries #1}
  }

\usepackage[authoryear]{natbib}
\title{Evolution and Asteroseismology of Pulsating Low-Mass White Dwarfs} 

\shorttitle{Evolution and Asteroseismology of Pulsating LMWDs
} 

\shortauthors{Calcaferro, C\'orsico, Althaus, Romero and Kepler} 

\author{
  \authorstyle{L.~M.~Calcaferro,$^{1,2}$ A.~H.~C\'orsico,$^{1,2}$, L.~G.~Althaus$^{1,2}$, A.~D. Romero$^3$ and S.~O. Kepler$^3$ }
	\newline\newline 
	$^1$\institution{Facultad de Ciencias Astron\'omicas y Geof\'isicas, Universidad Nacional de La Plata, Paseo del Bosque s/n (1900) La Plata, Argentina}\\ 
        $^2$\institution{Instituto de Astrof\'isica La Plata, CONICET-UNLP, Paseo del Bosque s/n, 1900, La Plata, Argentina}\\ 
	$^3$\institution{Departamento de Astronomia, Universidade Federal do Rio Grande do Sul, 
          Av. Bento Goncalves 9500, Porto Alegre 91501-970, RS, Brazil}\\
        \email{lcalcaferro,acorsico,althaus@fcaglp.unlp.edu.ar; alejandra.romero@ufrgs.br,kepler@if.ufrgs.br} 
      }

\begin{document}

\maketitle 

\thispagestyle{firstpage} 


\newabstract{
  Many low-mass white dwarfs are being discovered in the field of
  our galaxy and some of them exhibit $g$-mode pulsations,
  comprising the extremely low-mass variable (ELMV) stars class. Despite
  it is generally believed that these stars are characterized by thick
  H envelopes, from stellar evolution considerations, the existence of
  low-mass WDs with thin H envelopes is also possible.
  We have performed detailed asteroseismological fits to all the known
  ELMVs to search for a representative model by employing a
  set of fully evolutionary models that are representative of low-mass
  He-core white dwarf stars with a range of stellar masses
  $[0.1554-0.4352]\ M_{\odot}$, effective temperatures $[6000-10000]\ $K,
  and also with a range of H envelope thicknesses $-5.8 \lesssim \log(M_{\rm H}/M_{\star}) \lesssim -1.7$, hence expanding the space of parameters.  We found that some of the stars under analysis are characterized by thick H envelopes, but others are better represented by models with thin H envelope .}


\section{Introduction}
Low-mass white dwarfs (LMWDs), which are characterized by $M_{\star} \lesssim 0.45 M_{\odot}$, are thought to be formed by strong mass-loss episodes at the red giant branch (RGB) of low-mass stars in binary systems before the occurrence of the He flash, so they are expected to harbor He cores \citep[see][for instance]{2013A&A...557A..19A,2016A&A...595A..35I}. Among them, there is a population of WDs with very low mass: ELMs with $M_{\star} \lesssim 0.18-0.20 M_{\odot}$, characterized by $5 \lesssim \log(g) \lesssim 7$ and $8\,000 \lesssim T_{\rm eff} \lesssim 22\,000\ $K. For stars with masses greater than $0.18-0.20 M_{\odot}$ the WD progenitors are expected to experience CNO flashes that reduce their hydrogen content, making them unable to sustain nuclear burning. Then, they are expected to have shorter evolutionary timescales in comparison with stars with masses lower than $0.18-0.20 M_{\odot}$, whose progenitors are not expected to experience H flashes, and hence, end up with thicker H envelopes. Consequently, they sustain residual nuclear burning with the resulting longer evolutionary timescales. The mentioned upper-mass limit for ELM WDs is not only motivated by these physical differences but also because they differ in their pulsational properties \citep{2013A&A...557A..19A,2014A&A...569A.106C}, however this value depends on the metallicity of the WD progenitors \citep[see][for instance]{2016A&A...595A..35I}.

These stars are detected by the ELM, SPY, WASP and SDSS surveys, among others, and some of them show $g$-mode pulsation periods \citep{2012ApJ...750L..28H,2013ApJ...765..102H,2013MNRAS.436.3573H,2015MNRAS.446L..26K,2015ASPC..493..217B,2017ApJ...835..180B,2018MNRAS.478..867P,2018arXiv180511129B}, allowing the study of their interiors by applying the tools of WD asteroseismology. In the theoretical plane, adiabatic pulsational analyses show that $g$ modes in ELMVs with $M_{\star} < 0.18 M_{\odot}$ are restricted mainly to the core regions \citep{2010ApJ...718..441S,2012A&A...547A..96C,2014A&A...569A.106C}. Then, we can constrain the core chemical structure of these stars. Nonadiabatic stability computations \citep{2012A&A...547A..96C,2013ApJ...762...57V,2016A&A...585A...1C} predict that there are unstable $g$ and $p$ modes excited by $\kappa-\gamma$ \citep{Unno89}  mechanism acting at the H-ionization zone. Also, the $\varepsilon$ mechanism may destabilize some short period $g$ modes \citep{2014ApJ...793L..17C}.

The asteroseismological techniques have been successfully applied to WD stars \citep{WK08,FB08,2010A&ARv..18..471A}. One of the main asteroseismological avenues, developed at La Plata Observatory, involves the calculation of fully evolutionary models characterized by chemical profiles resulting from all the processes experienced during the evolution of the WD progenitors, and this is the approach we follow, that has already been employed in several cases; see, for instance, the cases for GW Virginis stars \citep{2007A&A...461.1095C,2007A&A...475..619C,2008A&A...478..869C,2009A&A...499..257C,2014MNRAS.442.2278K,2016A&A...589A..40C}, DBV stars \citep{2012A&A...541A..42C,2014A&A...570A.116B} and ZZ Ceti stars \citep{2012ApJ...757..177K,2012MNRAS.420.1462R,2013ApJ...779...58R,2017ApJ...851...60R}.

In \cite{2017A&A...607A..33C} we have performed the first asteroseismological analysis of all the known ELMVs by computing period-to-period fits employing radial and non-radial $g-$ and $p-$mode pulsation periods of low-mass He-core WD evolutionary models with stellar masses between $0.1554$ and $0.4352 M_{\odot}$, resulting from the computations of \cite{2013A&A...557A..19A}, that take into account the binary evolution of the progenitor stars. In that work, we were able to find solutions in most of the cases but in all of them, there were multiple possible solutions. In addition, for most of the stars the derived asteroseismological models are more massive in comparison with the spectroscopic determinations, something that could be related to the fact that the only considered low-mass He-core WD models are characterized by outer H envelopes coming from the stable mass loss scenario via Roche-lobe overflow. It cannot be discarded, however, that some of these stars can have thinner H envelopes that could result from common-envelope evolution of close binary systems \citep{2016MNRAS.460.3992N,2016MNRAS.462..362I,2017MNRAS.470.1788C}, or from the lost of the envelope of a RGB star induced by an inspiralling giant planet \citep{1998A&A...335L..85N,2002PASP..114..602D,2017arXiv170608897S}, although this issue is currently under debate. Then, by virtue of the mentioned considerations, we expand the space of parameters by introducing the H envelope thickness as an additional adjustable model parameter and employing this new grid, we perform a thorough analysis by considering not only a range in stellar mass ($0.1554 < M_{\star} < 0.4352 \ M_{\odot}$) and effective temperature ($13000 \gtrsim T_{\rm eff} \gtrsim 6000\ $ K) but also, in the H-envelope thickness ($-5.8 \lesssim \log(M_{\rm H}/M_{\star}) \lesssim -1.7$, depending on the stellar mass). Thus, our fits are done over $17\, 000$ WD configurations.

\section{Methods}
For the present analysis, we employed the realistic configurations for low-mass He-core WD stars computed by \cite{2013A&A...557A..19A}, that imitates the binary evolution of the progenitor stars assuming initial configurations consisting of a $1.0 M_{\odot}$ Main Sequence (donor) star and a $1.4 M_{\odot}$ neutron star companion as the other component. We refer the reader to that work for details of the physics and the numerical code (LPCODE) employed. When the initial orbital period is varied (between $0.9$ and $300 $d) different WD models are obtained (with $M_{\star} $= $0.1554$, $0.1612$, $0.1650$, $0.1706$, $0.1762$, $0.1805$, $0.1863$, $0.1921$, $0.2025$, $0.2390$, $0.2707$, $0.3205$, $0.3624$ and $0.4352 M_{\odot}$). Adiabatic pulsation periods for non-radial $g$ modes with $\ell= 1$ and $\ell=2$ were taken from \cite{2014A&A...569A.106C} in the case of WD models with canonical H envelope thicknesses. For WD models having thinner H envelopes, we computed the periods for the present work. In both cases, the pulsation periods were computed employing the adiabatic version of the LP-PUL pulsation code \cite{2006A&A...454..863C}.

\begin{figure}[t]
  \centerline{\includegraphics[width=1.0\columnwidth]{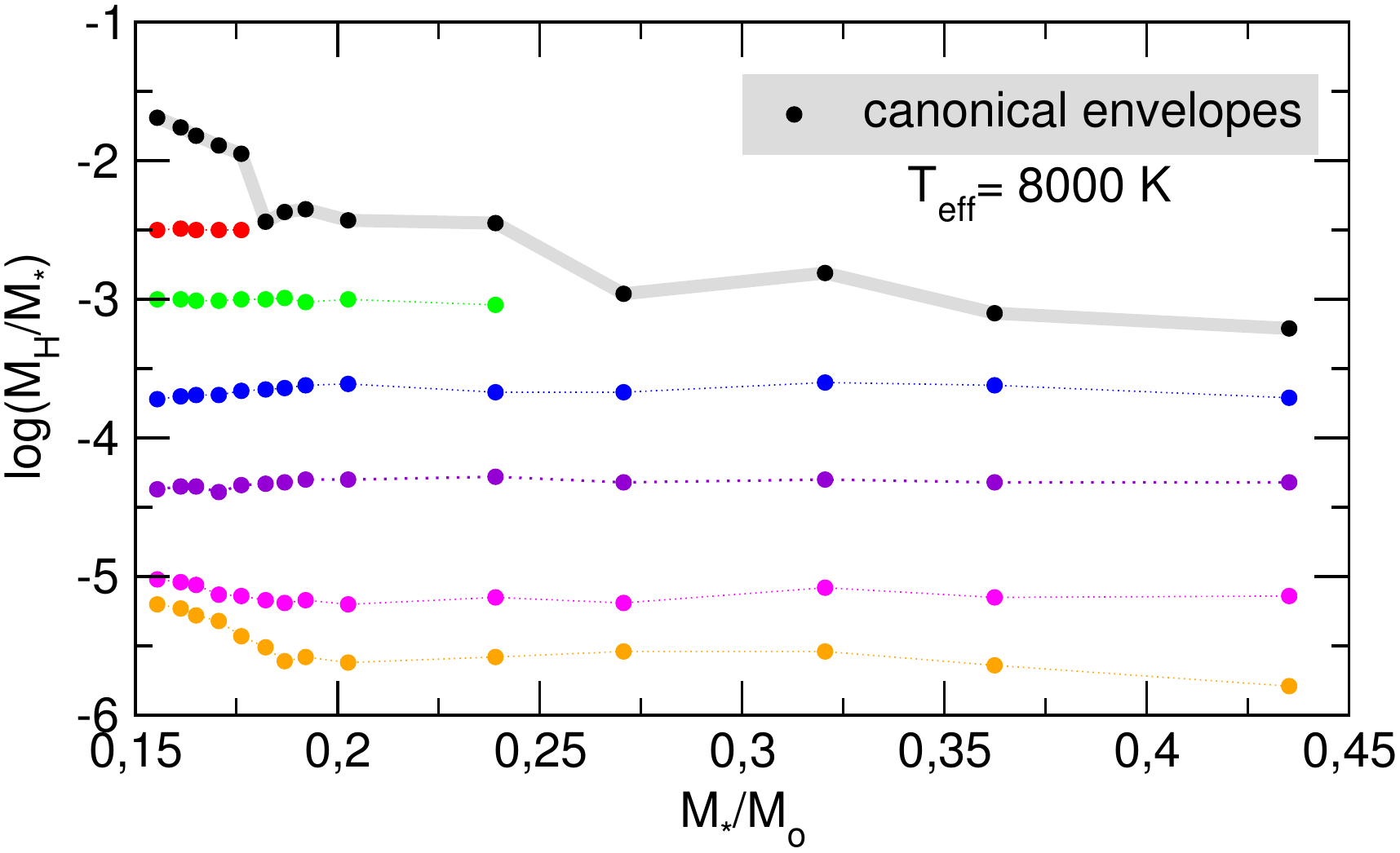}}
  \caption{Grid of low-mass He-core WD evolutionary sequences considered in this work represented in the $M_{\star}-\log(M_{\rm H}/M_{\star})$ plane. Small circles indicate a sequence of WD models with specific values of thickness of the H envelope and stellar mass at $T_{\rm eff} \sim 8000\ $K. The gray line connects circles corresponding to the values of the maximum H envelope thickness as predicted by the evolutionary computations of \cite{2013A&A...557A..19A}.} 
  \label{Henvelopes}
\end{figure}

\begin{figure} 
\begin{center}
\includegraphics[clip,width=8 cm]{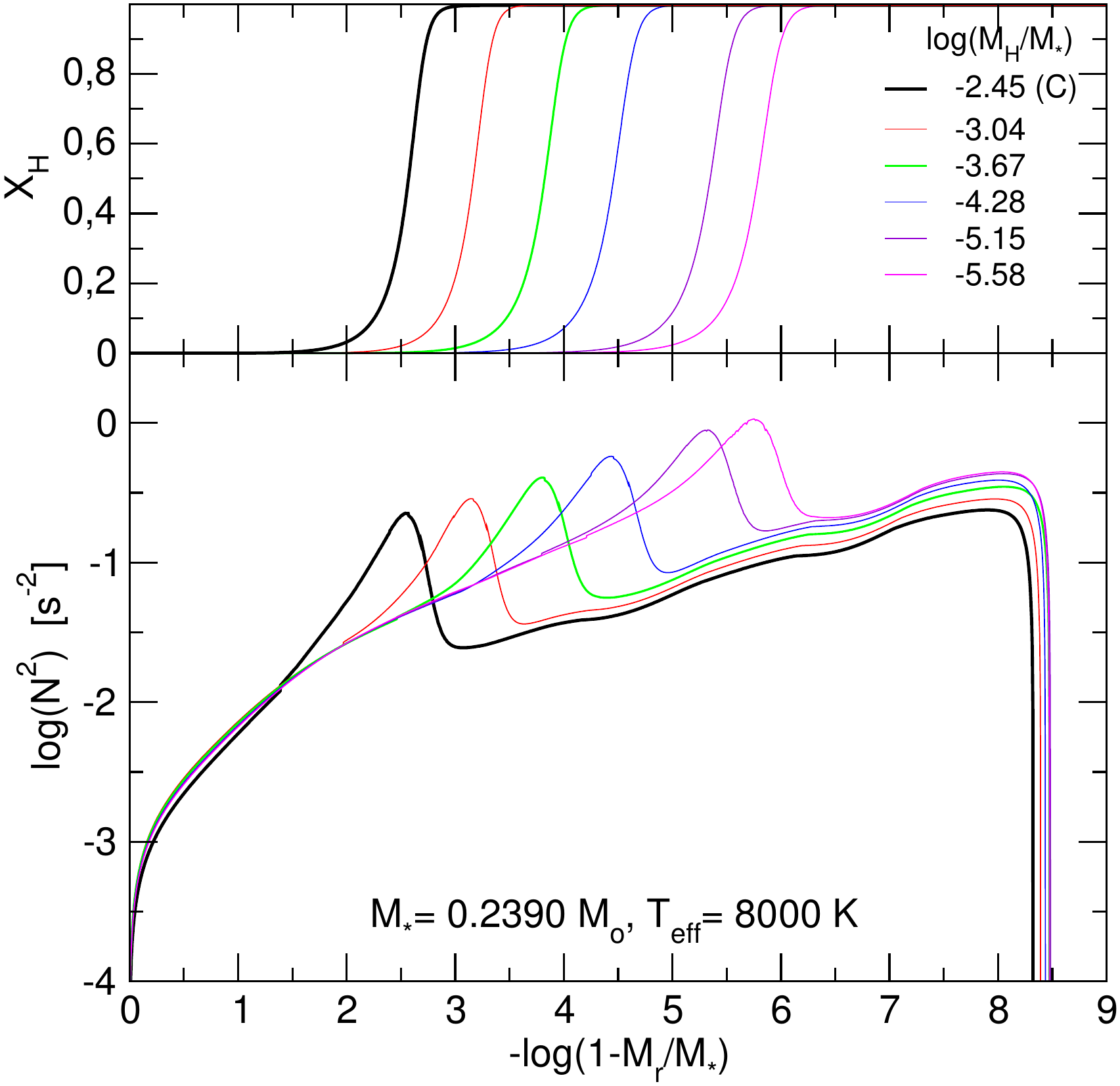} 
\caption{Upper panel: chemical profiles of H for WD models with $M_{\star}= 0.2390 M_{\odot}$ at $T_{\rm eff} \sim 8000$ K and different thicknesses of the H envelope (the black line indicates the canonical envelope). Lower panel: run of the logarithm of the squared Brunt–V\"ais\"al\"a frequency for each model.} 
\label{profiles} 
\end{center}
\end{figure}

To generate this new set of sequences with different H envelope thicknesses, for each sequence characterized by a given value of $M_{\star}$ and a thick (canonical) value of $M_{\rm H}$ we have artificially replaced $^1$H by $^4$He from a given mesh point in order to obtain certain values of the H envelope thickness. This is done at very high $T_{\rm eff}$ values at the final cooling track to ensure that any nonphysical transitory effects associated with this procedure have concluded long before the models get to the pulsating stage of ELMV WD stars. Time-dependent element diffusion was allowed to act after implementing the change in the thicknesses of the H envelope. Diffusion erodes considerably the chemical profiles at the transition regions. In Fig.~\ref{Henvelopes} we show a graphical representation of the values that result for the different H envelope thicknesses, for every stellar mass under consideration, at $T_{\rm eff} \sim 8000$ K. A gray line connects the canonical values of the H-envelope mass, as stellar evolution predicts. In the upper panel of Fig.~\ref{profiles}, we display the internal chemical profiles for H corresponding to WD models at $T_{\rm eff} \sim 8000$ K with $M_{\star}= 0.2390 M_{\odot}$, where the black line represents the profile corresponding to the canonical envelope and the lines of different colors correspond to the thin H envelopes. In all the envelopes of these models, the He/H transition region has a single-layered shape. The  shape of the chemical profiles leaves significant signatures in the run of the squared critical frequencies, and in particular, in the Brunt-V\"ais\"al\"a frequency ($N$). In the lower panel of Fig.~\ref{profiles}, we display the logarithm of the squared Brunt–V\"ais\"al\"a frequency, where we can see the clear connection between the chemical transition regions (upper panels) and the features in the run of the Brunt–V\"ais\"al\"a frequency for each model.

\section{Results}

\begin{figure}[t]
  \centerline{\includegraphics[width=1.0\columnwidth]{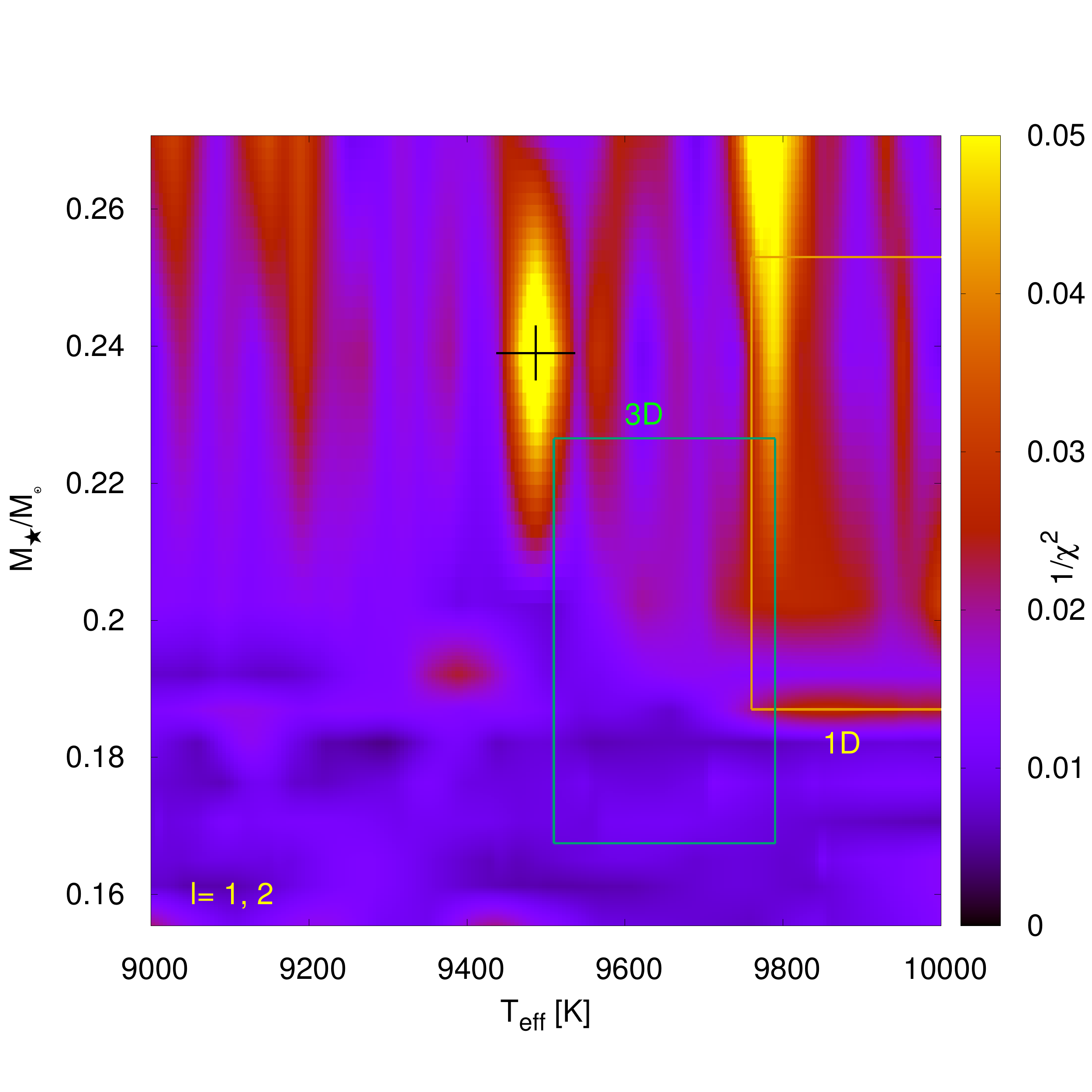}}
  \caption{Projection on the $T_{\rm   eff}$ vs $M_{\star}/M_{\star}$ plane of the inverse of the quality function for $\ell= 1, 2$ $g$ modes, considering the set of observed periods for SDSS J151826.68+065813.2. The value of the thickness of the H envelope for each stellar mass corresponds to the sequence with the largest value of the inverse of the quality function for that stellar mass. The boxes depict the spectroscopic parameters for this star, along with their uncertainties, for the 1D and 3D model atmosphere. These spectroscopic boxes are defined considering $\pm \sigma$. The ranges in the three axes are focused on values of interest.} 
  \label{j1518}
\end{figure}

\begin{figure*}[t]
  \begin{center}
    \subfigure[Set with $\Pi_i^{\rm  O} - \sigma$]{\label{j1840-sigma}\includegraphics[clip,width=5.2 cm]{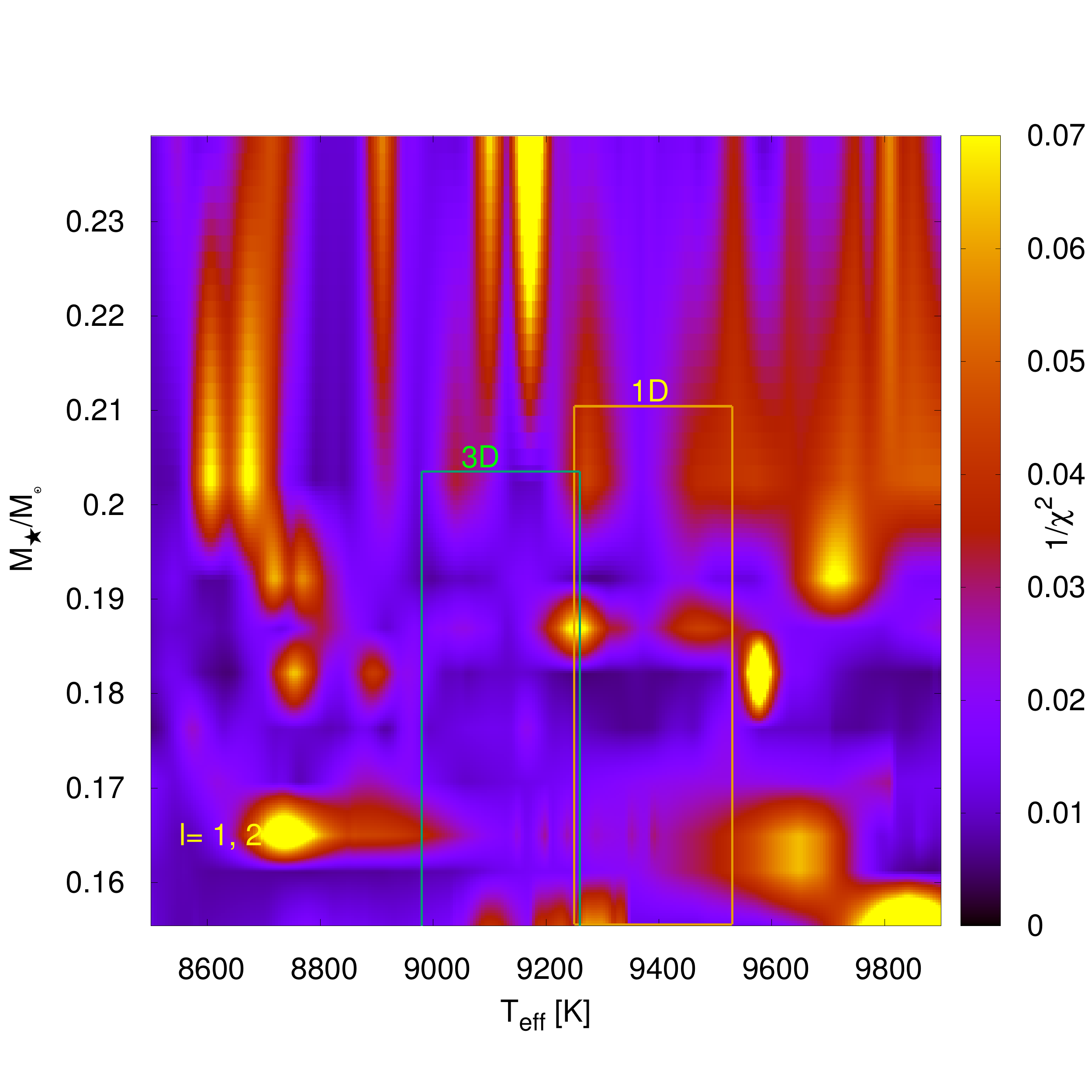}}
    \subfigure[Set with $\Pi_i^{\rm  O}$]{\label{j1840-center}\includegraphics[clip,width=5.2 cm]{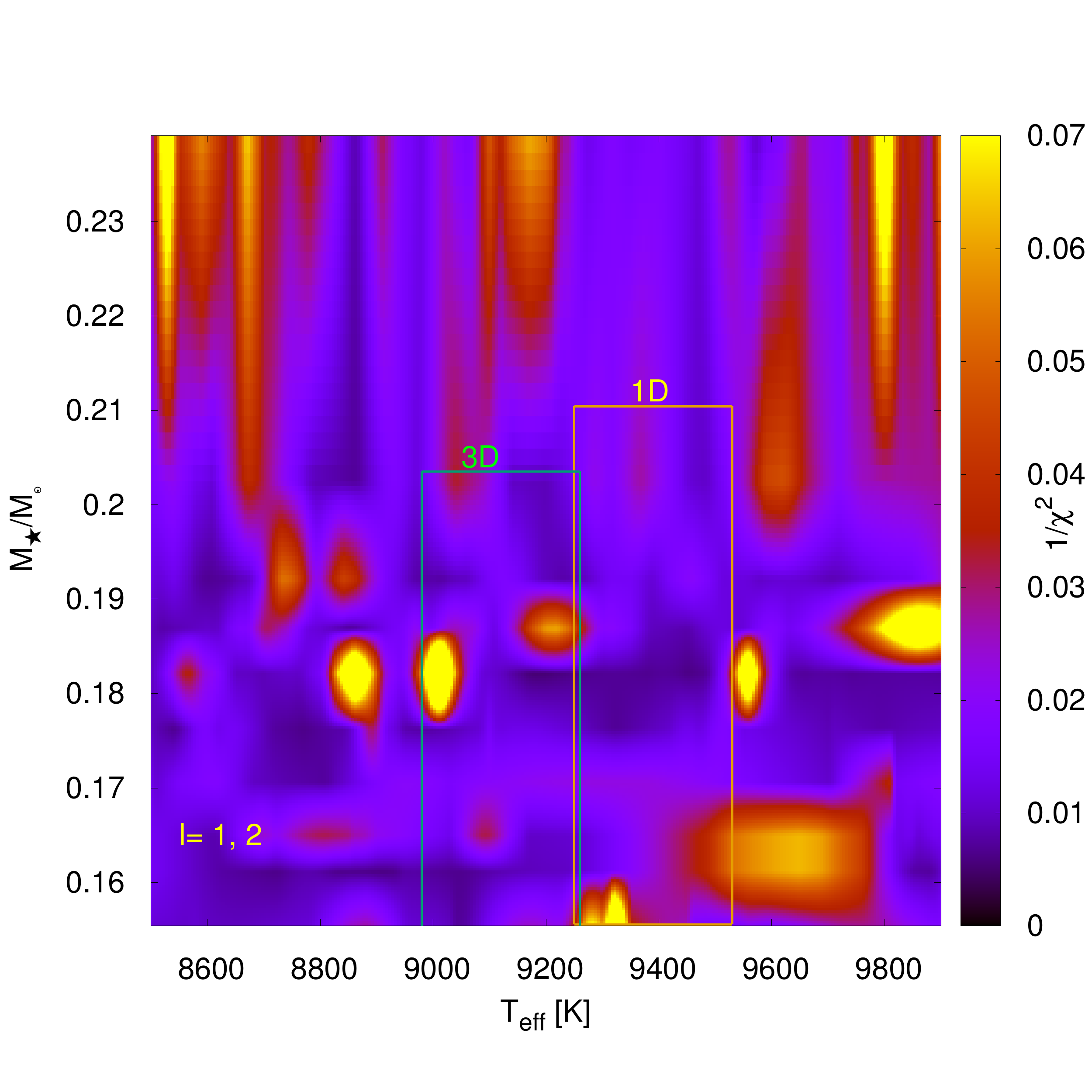}} 
    \subfigure[Set with $\Pi_i^{\rm  O} + \sigma$]{\label{j1840+sigma}\includegraphics[clip,width=5.2 cm]{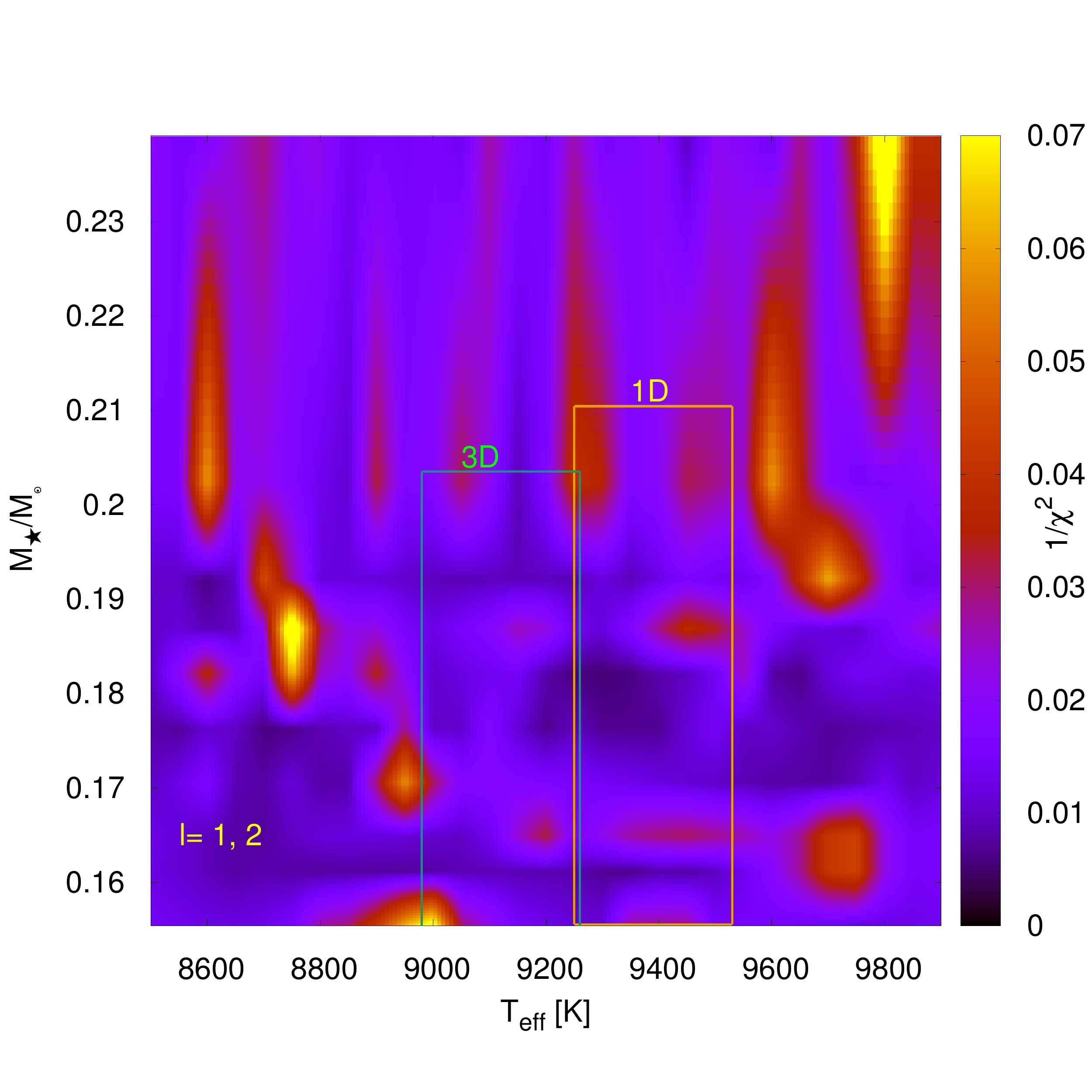}}
  \end{center}
  \caption{Same as ~\ref{j1518} but for the three different set of  periods of SDSS J184037.78+642312.3, considering no uncertainty for all the periods (middle panel), and the sets with $- \sigma$ (left panel) and $+ \sigma$ (right panel) only for $\Pi_i^{\rm  O}= 3930.0\ $s.}  
\label{j1840}
\end{figure*}

With the aim of finding an asteroseismological model whose periods best match the observed periods of every ELMV, we asses the quality function given by:

\begin{equation}
\label{chi}
\chi^2(M_{\star},  T_{\rm   eff}, M_{\rm H})=   \frac{1}{n} \sum_{i=1}^{n}   \min[(\Pi_i^{\rm   O}-   \Pi_k^{\rm  T})^2], 
\end{equation}

\noindent being $n$ the number of observed periods. The ELM model with the lowest value of $\chi^2$, if exists, is adopted as the ``best-fit model''.  We compute this merit function $\chi^2=\chi^2(M_{\star}, T_{\rm eff}, M_{\rm H})$ for our set of stellar masses, covering a wide range in effective temperature $13000 \gtrsim T_{\rm eff} \gtrsim 6000\ $ K and also considering the thickness of the H envelope in the interval $-5.8 \lesssim \log(M_{\rm H}/M_{\star}) \lesssim -1.7$. Firstly, we consider that all of the observed periods ($\Pi_i^{\rm  O}$) for each ELMV are associated with $\ell= 1$ $g$ modes and we asses the quality function given by Eq.~(\ref{chi}). Next, we consider a mix of $g$ modes associated with both $\ell= 1$ and $\ell=2$. Given that the solutions we obtain are more appropriate for the latter, in this work we only show those cases. In Figures~\ref{j1518} and \ref{j1840} we show the projection on the effective temperature versus the stellar mass plane of the inverse of the quality function, $(\chi^2)^{-1}$, for the ELMV under consideration, taking the corresponding set of observed periods into account, in analogy with \cite{2017A&A...607A..33C}. In these figures we include $T_{\rm eff}$ and $M_{\star}$ of the target star, along with their uncertainties for the 1D (orange box) and 3D \citep[][green box]{2015ApJ...809..148T} model atmosphere  determinations.  For all stellar masses, we considered an uncertainty of $15\%$ of the total mass, which is the characteristic difference in the value of the mass as derived from independent sets of evolutionary tracks \citep{2017A&A...607A..33C}. Each point in the maps corresponds to an H envelope mass value ($M_{\rm H}/M_{\star}$) that maximizes the value of $(\chi^2)^{-1}$ for that stellar mass and effective temperature. If there is a single maximum for a given star, we adopt the corresponding model as the asteroseismological solution. However, as in the cases under analysis there are multiple possible solutions, we need to apply an external constraint, i.e. the uncertainty in the effective temperature given by the spectroscopy and, at variance with \cite{2017A&A...607A..33C}, we additionally employ the constraint of the stellar mass as given by the spectroscopic determinations.

We applied the mentioned procedure to all the known (and suspected) ELMVs, but here, we only show the results for two of them as an example. In Fig.~\ref{j1518} we show the case for SDSS J151826.68+065813.2 (J1518, for short). According to the 1D model atmosphere, this star is characterized by $T_{\rm eff}= 9990 \pm 140\ $K, $\log(g)= 6.80 \pm 0.05\ $ [cgs] and $M_{\star}= 0.220 \ M_{\odot}$ \citep{2013ApJ...765..102H}, and for the 3D model, $T_{\rm eff}= 9650 \pm 140\ $K, $\log(g)= 6.68 \pm 0.05\ $[cgs] and $M_{\star}= 0.197 \ M_{\odot}$ \citep{2015ApJ...809..148T}). The seven periods observed for this star are $\Pi_i^{\rm  O}= 1335.318 \pm 0.003$, $1956.361 \pm 0.003$, $2134.027 \pm 0.004$, $2268.203 \pm 0.004$, $2714.306 \pm 0.003$, $2799.087 \pm 0.005$ and $3848.201 \pm 0.009\ $s, according to \cite{2013ApJ...765..102H}. As we can see in the figure, there are not any
solutions within the spectroscopic boxes, however there is a possible solution 
at $\sim 9487\ $K, characterized by $0.2390 \ M_{\odot}$, $\log(M_{\rm H}/M_{\star})= -3.67$ and $(\chi^2)^{-1}= 0.07$. This is the best period fit in the considered ranges, and it lies closely to the spectroscopic parameters, then we may adopt it as a solution for J1518. Once a model is adopted, it is worth determining the difference between the observed and theoretical periods. In this way, we assess the absolute period differences defined as $|\delta\Pi|= |\Pi^{\rm O}-\Pi^{\rm T}|$ and we show in Table~\ref{tab:perj1518} the results for this case. In column 6 we also indicate the value of the linear non-adiabatic growth rate, $\eta$ ($\eta \equiv-\Im(\sigma)/ \Re(\sigma)$, being $\Re(\sigma)$ and $\Im(\sigma)$ the real and the imaginary part, respectively, of the complex eigenfrequency $\sigma$ computed  with the non-adiabatic version of the {\tt LP-PUL} pulsation code  \citep{2006A&A...458..259C,2016A&A...585A...1C}). If $\eta$ is positive (negative), the mode is unstable (stable). From this table we can see that most of the periods corresponding to the adopted asteroseismological model are associated with pulsationally unstable modes.

In Fig.~\ref{j1840-center} we show the case for SDSS J184037.78+642312.3 (J1840, for short), with spectroscopic parameters $T_{\rm eff}= 9390 \pm 140\ $K, $\log(g)= 6.49 \pm 0.06\ $ [cgs] and $M_{\star}= 0.183\ M_{\odot}$ for the 1D model atmosphere \citep{2012ApJ...750L..28H}, and $T_{\rm  eff}= 9120 \pm 140\ $K, $\log(g)= 6.34 \pm 0.05\ $[cgs] and $M_{\star}= 0.177 \ M_{\odot}$, for the 3D model atmosphere \citep{2015ApJ...809..148T}. We consider the set of the five observed periods ($\Pi_i^{\rm  O}= 1164.15 \pm 0.38$, $1578.7 \pm 0.65$, $2376.07 \pm 0.74$, $3930.0 \pm 300$ and $4445.3 \pm 2.4\ $s) according to \cite{2012ApJ...750L..28H}. 
The figure shows the existence of multiple possible solutions, however there is a very good solution that lies within the 3D model atmosphere at $\sim 9007\ $K, for $0.1805 \ M_{\odot}$ and $\log(M_{\rm H}/M_{\star})= -2.44$ (i.e., a model with a canonical envelope), with $(\chi^2)^{-1}= 0.21$. The fact that one of the observed periods has a meaningful uncertainty ($\Pi_i^{\rm  O}= 3930.0\ $s, $\sigma= 300\ $s) led us to repeat the analysis but considering a set with $-\sigma$ and another one with $+\sigma$ and we show the results in Figs.\ref{j1840-sigma} and \ref{j1840+sigma} We found that solutions change considerably when varying that period and, as a result, we were not able to find a unique solution, but a range of possible solutions with parameters between $M_{\star}= 0.1554 - 0.1869\ M_{\odot}$, $T_{\rm eff} \sim 8997 - 9244\ $K, and $M_{\rm H}/M_{\star}= 6.53 \times 10^{-6} - 2.04 \times 10^{-2}$. 

\begin{table*}[t]
\centering
\caption{Observed and theoretical periods  ($\ell= 1, 2$)  for the asteroseismological model for J1518 with $M_{\star}= 0.2390\ M_{\odot}$, $T_{\rm eff}\sim 9487\ $ K and  $\log(M_{\rm H}/M_{\star})= -3.67$. The harmonic degree $\ell$, the radial order $k$, the absolute period difference, and the non-adiabatic growth rate for each theoretical period are also displayed.}
\begin{tabular}{cccccccc}
\hline
\hline
 $\Pi^{\rm O}$[s] & $\Pi^{\rm T}$[s] &  $\ell$ & $k$ & $|\delta\Pi|$[s] & $\eta[10^{-5}]$ &
Remark\\
\noalign{\smallskip}
\hline
$1335.318$& $1329.599$& $2$ & $28$&  $5.719$ & $0.463$ & unstable \\
$1956.361$& $1959.913$& $1$ & $24$&  $3.552$ & $0.653$ & unstable \\
$2134.027$& $2131.306$& $2$ & $46$&  $2.721$ & $0.504$ & unstable \\
$2268.203$& $2266.188$& $1$ & $28$&  $2.015$ & $0.766$ & unstable \\
$2714.306$& $2717.686$& $2$ & $59$&  $3.380$ & $-0.373$ & stable \\
$2799.087$& $2802.873$& $1$ & $35$&  $3.786$ & $1.14$ & unstable \\
$3848.201$& $3851.967$& $2$ & $84$&  $3.766$ & $-4.96$ & stable \\
\hline
\end{tabular}
\label{tab:perj1518}
\end{table*}

\vspace*{-0.5em}
\section{Conclusions}

In this work we have presented an asteroseismological analysis carried out on pulsating ELM WD stars on the basis of our complete set of fully evolutionary models representative of low-mass He-core WDs with a range of H envelope thicknesses. We generated a new grid of models for every stellar mass in our set and then, we performed an asteroseismological analysis to all the known (and suspected) ELMV stars, as in \cite{2017A&A...607A..33C}, but employing this larger set of evolutionary sequences instead, that expands the parameter space by incorporating the thickness of the H envelope as a free parameter.
We found multiplicity of solutions in all the cases, that may be due to the few periods detected in these stars. Only with the inclusion of external constraints (i.e., spectroscopic parameters) we were able to adopt a model, however in three cases, we could only indicate an interval of possible solutions. These results are summarized in Table~\ref{tab:tablafinal}. Also, some datasets exhibit one or more periods with significant uncertainty ($\sigma$), then we carried the same analysis out but considering $\pm \sigma$, for the most uncertain period. We found a meaningful variation in the results, one of these cases being J1840. Furthermore, some of the solutions are characterized by thick (canonical) H envelopes and some by thin H envelopes. This reinforces the findings of \cite{2018A&A...614A..49C} about the possible existence of ELM WDs with thin H envelope, hence leading to the possibility that they could have been formed through unstable mass loss, maybe via common-envelope episodes or from the lost of the envelope of a RGB star induced by an inspiralling giant planet.

Our results show that with the current amount of observed periods of all the ELMVs, it is not possible to find a unique solution compatible with the spectroscopic determinations. It becomes an even more difficult task when one (or more) periods have a large uncertainty. Considering that we are employing a complete set of fully evolutionary models representative of He-core ELM WDs, with different H-envelope thicknesses, we have now reached a limit regarding the possibility of the asteroseismology of adopting a model through the period fit on the basis of this grid, in order to determine the internal structure of these stars. This indicates the necessity of having richer observations of the pulsations of these stars, to be able to find more robust asteroseismological solutions. Finally, the discovery of new ELMVs is also a pressing need in order to have further knowledge of their internal structure, the nature of their progenitors, and the evolutionary channels that originates them  \citep[see][]{2018A&A...614A..49C}.

\begin{table*}[t]
\centering
\caption{Main features of the adopted asteroseismological models for the known (and suspected) ELMVs.}
\begin{tabular}{lcccccc}
\hline
\hline
Star & $T_{\rm eff}$ [K]  & $\log(g)$ [cgs]  &  $M_{\star}[M_{\odot}]$ &  $\log(M_{\rm H}/M_{\star})$ & $\log(R_{\star}/R_{\odot})$ & $\log(L_{\star}/L_{\odot})$ \\
\hline
J1840      &[8997,9244] &[5.8276,6.7524] & [0.1554,0.1869] &[-5.19,-1.69] &[-1.5216,-1.0992]  &[-2.2166,-1.4249]   \\
J1112$^{\rm *}$  & 9301  &     5.9695     & 0.1612          &-1.76          &-1.1623           &-1.4932 \\
J1518          & 9487  &     6.9994     & 0.2390          &-3.67           &-1.5916           &-2.3200 \\
J1614          & 8989  &     6.4468     & 0.1612          &-4.35           &-1.4009           &-2.0232  \\
J2228$^{\rm *}$ & 7710  &     6.1738      & 0.1554          &-1.69           & -1.2725         &-2.0409 \\
J1738      &[8883,9273] &[6.0506,6.6923] &[0.1612,0.1921] &[-5.43,-1.76]   &[-1.5057,-1.2029] &[-2.2548,-1.6560] \\
J1618      &[8919,9231] &[6.2661,6.7568] &[0.1650,0.1921] &[-5.06,-1.89]   &[-1.5178,-1.2982] &[-2.2447,-1.8401] \\
J1735$^{\rm *}$ & 8075  &     6.2241       & 0.1612         & -1.76         & -1.2899           &-1.9957 \\
J2139         & 8173  &     6.3355       & 0.1612         & 2.49          & -1.3453           &-2.0820 \\

\hline
\end{tabular}
\label{tab:tablafinal}
{\footnotesize
Note: $^{\rm *}$ Solution with canonical H envelope.}
\end{table*}


\bibliography{papers}

\end{document}